# Cryptic photosynthesis – extrasolar planetary oxygen without a surface biological signature

Charles S Cockell,

Centre for Earth, Planetary, Space and Astronomical Research, Open University, Walton Hall, Milton Keynes MK7 6AA, UK

Lisa Kaltenegger,

Harvard-Smithsonian Center for Astrophysics. 60 Garden St. MS20, Cambridge, MA 02138, USA

John A Raven,

Division of Plant Sciences, University of Dundee at SCRI, Scottish Crop Research Institute, Invergowrie, Dundee DD2 5DA, UK

On the Earth, photosynthetic organisms are responsible for the production of virtually all of the oxygen in the atmosphere. On the land, vegetation reflects in the visible, leading to a 'red edge' that developed about 450 Myr ago and has been proposed as a biosignature for life on extrasolar planets. However, in many regions of the Earth, and particularly where surface conditions are extreme, for example in hot and cold deserts, photosynthetic organisms can be driven into and under substrates where light is still sufficient for photosynthesis. These communities exhibit no detectable surface spectral signature to indicate life. The same is true of the assemblages of photosynthetic organisms at more than a few metres depth in water bodies. These communities are widespread and dominate local photosynthetic productivity. We review known cryptic photosynthetic communities and their productivity. We link geomicrobiology with observational astronomy by calculating the disk-averaged spectra of cryptic habitats and identifying detectable features on an exoplanet dominated by such a biota. The hypothetical cryptic photosynthesis worlds discussed here are Earth-analogs that show detectable atmospheric biomarkers like our own planet, but do not exhibit a discernable biological surface feature in the disc-averaged spectrum.

## INTRODUCTION

The determination of the presence of biological processes on extrasolar planets will depend on the detection of spectral signatures that indicate habitability (see e.g. Owen 1980; Léger *et al.* 1993; Des Marais *et al.*, 2002; Selsis *et al.*, 2002; Wolstencroft and Raven, 2002; Raven and Wolstencroft, 2004; Kaltenegger and Selsis, 2007; Segura *et al.*, 2007 for detailed discussion).

These spectral signatures broadly fall into two classes: 1) atmospheric spectra of major and trace gases that suggest gas exchange reactions out of equilibrium with purely geochemical processes (such as high simultaneous concentrations of $O_2$, $O_3$, $CH_4$ and $N_2O$), 2) spectra from surface reflections that indicate absorption from pigments or compounds that are not mineralogical and may be derived from biotic syntheses. Here we discuss how these spectral features might appear in disc-averaged spectra of an Earth-like planet observed from space and their detectability.

In the latter category of spectral signatures, the 'red edge', which is characteristic of terrestrial vegetation, has received considerable attention (e.g. Seager *et al.*, 1995, Tinetti et al 2006, Kiang et al 2007a). The red edge results from a strong chlorophyll absorption in the red region of the spectrum contrasted with a strong reflectance in near-infrared wavelength regions. Photosynthetic plants have strong reflectance, possibly as an evolved mechanism that restricts metabolically dangerous overheating in air. Direct selection for this red edge is certainly possible, and some desert plants have highly reflective surfaces with a reflectance of 0.05 – 0.25 between 400 and 700 nm, rising to as much as 0.80 at 750nm, and a reflectance of 0.35 – 0.80 and 0.05 – 0.57 from 750 to 1000 nm and from 1000 to 2000 nm respectively. These properties of desert plants restrict absorption or transmission of photosynthetically active radiation (PAR) as well as of longer wavelengths (Gates, 1980; Jones, 1992; Merzlyak *et al.*, 2002).

The vegetation red edge (VRE) on Earth is distinctive. It can be used to map vegetation with very high resolution data. Several groups have measured the integrated Earth spectrum with the technique of Earthshine, which uses sunlight reflected from the non-illuminated, or "dark", side of the moon. Earthshine measurements have shown that detection of Earth's red edge could be feasible but made very difficult owing to its broad, essentially featureless spectrum and the interfering effects of cloud coverage (e.g. Montanes-Rodriguez *et al.*, 2007; Turnbull *et al.*, 2005). Averaged over a spatially unresolved hemisphere of Earth, the additional reflectivity of this spectral feature is a few percent.

The exact wavelength and strength of the spectroscopic red edge depends on the plant species and environment (Wolstencroft and Raven, 2002; Tinetti *et al.*, 2006; Kiang *et al.*, 2007a; Stomp *et al.*, 2007, Arnold *et al.*, 2002, 2009). Kiang *et al.* (2007a) provide a synthesis of the maximum absorption wavelengths found in a diversity of photosynthetic organisms, including microbial phototrophs. Despite some variations between organisms, they suggest that general rules can be devised which can be used to predict where photosynthetic organisms are most likely to absorb in the spectrum (see also Stomp *et al.*, 2007). They also propose that these spectral signatures may shift according to the spectral output of the parent star (Kiang *et al.*, 2007b).

The presence of a red edge, or any surface biological signature, depends upon the photosynthetic organisms growing on the planetary surface influencing reflected light. In many environments, microbial phototrophs are hidden beneath substrates, but they remain sufficiently close to the surface to collect light for photosynthesis (Cockell and Raven, 2004). This phenomenon (which can be called 'cryptic photosynthesis') is particularly common in extreme environments where organisms may escape detrimental conditions on the surface, such as desiccation and UV radiation, within shielded micro-habitats. This phenomenon raises the possibility of a 'false-negative' detection of life if the alteration of surface reflection spectra, including the red edge, is used as a criterion for habitability, i.e., an atmospheric spectral signature of photosynthesis – oxygen or ozone – indicates habitability but no surface reflected signature is found.

In this paper, we discuss the case of cryptic photosynthesis worlds. We combine geomicrobiology with observational astronomy to: 1) provide an overview of cryptic photosynthetic communities on the Earth, and 2) calculate the expected spectral signature from planets where a large proportion or all of its photosynthesis is cryptic.

**THE FORMS OF CRYPTIC PHOTOSYNTHESIS ON THE EARTH**
A number of regions of the Earth exhibit cryptic photosynthetic growth. Table 1 summarises the categories of cryptic photosynthesis that we recognise on the Earth with corresponding comments about their substrate requirements, what surface planetary geology would be needed to realise these requirements elsewhere (and thus how common the communities could be) and how productive some of these communities are.

*Rock and mineral-based cryptic photosynthesis*

**Chasmoendoliths**. The most ubiquitous habitat for cryptic photosynthesis, which makes the fewest assumptions about the host material, is the chasmoendolithic habitat, in which microorganisms inhabit fractures and cracks within rocks that are connected to the surface. Broady (1981) described a diversity of photosynthetic organisms inhabiting Antarctic rocks, and lichens have been described in Antarctic granites (De los Rios *et al.*, 2005). However, any process that produces cracks within rocks, e.g., aqueous and Aeolian weathering, heat fracturing from volcanic activity, freeze-thaw, and tectonic alteration, for example, provides habitats in which photosynthetic organisms can live, and in potentially any environment. Impact cratering is one universal process that fractures rocks and generates space for colonization by chasmoendoliths (Fig.1c) (Cockell *et al.*, 2005). Chasmoendoliths can therefore inhabit any type of rock, including igneous, metamorphic and sedimentary and therefore make the fewest assumptions about the geological conditions to be found elsewhere.

**Cryptoendoliths**. Cryptoendolithic communities are organisms growing within rock structural cavities and inter-grain spaces rather than macroscopic cracks connected to the surface as in the case of chasmoendoliths. They are therefore associated with permeable rocks and depend upon the transmission of light into the rock interstices (Nienow et al., 1988). Like hypoliths, they are predominantly found in sedimentary lithologies such as quartzitic sandstones (Fig.1b) (Friedmann, 1980, 1982; Saiz-Jimenez *et al.*, 1990), but they are also found in volcanic glass (Herrera *et al.*, 2009). Thus, although cryptoendolithic colonization requires more specific geological conditions than for chasmoendoliths they can be widespread in diverse lithologies. The potential habitats in which cryptoendoliths persist can be expanded into usually non-porous rocks by physical processes that systemically fracture and increase their permeability and translucence. One such process is impact cratering, which has been reported to transform gneiss, which is usually a poor substrate for cryptoendolithic communities, into a more porous and translucent material (Cockell *et al.*, 2002).

**Hypoliths**. Most hypoliths are to be found under quartz rocks, since quartz is translucent (Warren-Rhodes et al., 2006), allowing photosynthetic organisms to inhabit the underside and still receive enough light for growth. Raven *et al.* estimated the minimum light level required for photosynthesis to be 0.1 μmoles/m$^2$/s (Raven *et al.*, 2000). The reduction of visible light to approximately 0.06% of ambient (assuming it to be ~2,000 μmoles/m$^2$/s at midday) would be sufficient to extinguish photosynthesis, consistent with the observations reported for the colonization of quartz rocks. For example, at a depth of 40 mm under typical quartz rock collected from the Negev Desert, less than 0.01% of incident light (Berner and Evenari, 1978) penetrates while in colonized quartz rock from the Mojava desert less than 0.08% penetrates to below 25 mm (Schlesinger *et al.*, 2003). Although quartz is a common substrate, the potential range of hypolithic growth is vastly expanded by the process of freeze-thaw, which moves rocks, generating 'patterned' or 'sorted' ground (Fig. 1a). Small movements in the rocks generate spaces into which light can penetrate, accounting for the widespread colonisation of polar desert rocks by hypolithic communities (Cockell and Stokes, 2004, 2006). Polar deserts account for about $5 \times 10^6$ km$^2$ of Earth's surface (about 4% of the Earth's surface), but there is no reason why a planetary surface could not be entirely covered by polar desert subjected to freeze-thaw. The important attribute of the habitat is that the penetration of light to the underside of the rock is made possible by a ubiquitous physical process in polar regions (freeze-thaw) and it imposes no special geological requirements on the substrate, other than that it is not toxic to the biota growing beneath it.

**Euendoliths**. A more specialist form of endolithic growth is euendolithic (rock-boring) behaviour. The evolutionary selection pressures for this mode of rock colonization have been discussed (Cockell and Herrera, 2008). They may include selection pressures to acquire nutrients from rocks or benefits conferred by escaping extreme environmental conditions on the rock surface. Euendolithic growth is widespread in carbonate-dwelling fungi and cyanobacteria (Golubic *et al*., 1981; Hoppert *et al*., 2004; Smits, 2006). The activities of euendoliths play an important role in rock weathering in coastal areas (Schneider and Le Campion-Alsumard, 1999), where the majority of them are to be found, although some deserts hosts euendoliths (Bungartz and Wirth, 2007). Reports of euendolithic borings in deep-sea volcanic rocks have been made (e.g., Fisk *et al*., 1998), although the latter borings have not been associated with phototrophs. The euendolithic habit shows that low permeability and porosity does not preclude cryptic photosynthetic growth and that many rocks are susceptible to the innovation of active dissolution mechanisms that allow organisms to achieve cryptic growth within the rock substrate.

**Sands and desert crusts.** Many sands and soils harbour microbial crusts composed of diverse assemblages/communities of organisms with a photosynthetic component (Belnap and Lange, 2001). Many of these are visible from the surface, however some, particularly in extreme deserts, are located beneath the surface. Soil crusts have previously been recognised as potential analogs for life on the early Earth (Campbell, 1979). Cryptic photosynthesis within soils and sands (including desert crusts) depends upon the presence of translucent substrates which allow sufficient light penetration for photosynthetic growth in the grain interstices (Chen *et al*., 2007). The presence of granular materials such as sands and soils into which photosynthetic organisms can grow is a reasonable assumption for any planet with water and thus a hydrological cycle driving weathering processes, which produce granular material from parent minerals. The organisms that inhabit soil crusts are diverse and influenced by the specific mineralogy of a given crust, such as the diverse soil and desert crusts of the Colorado Plateau, USA (Garcia-Pichel *et al*., 2001).

Some of these communities grow in habitats with negligible precipitation, relying on 'occult precipitation' from dew and fog, and even atmospheres of very high relative humidity, to permit photosynthesis and growth (Lange *et al*., 1994). The organisms are generally desiccation tolerant (unlike the vegetative stages of most vascular land plants) and are poikilohydric (cannot control their water loss, as can the homoiohydric vascular land plants; Raven, 1995). Interestingly, rehydration after desiccation to a water content permitting growth is possible using water from a high atmospheric relative humidity alone with green algae (free-living and lichenised), but not with free-living cyanobacteria and cyanolichens (Lange *et al.,* 1992, 1994, 2001). It is likely that the extent of hydration is an important factor for many non-aquatic communities relying on cryptic photosynthesis (including all types of endoliths); lack of hydration limits photosynthesis by inhibiting metabolism, while excessive hydration can restrict gas exchange with atmosphere as a result of the much lower diffusion constants of $CO_2$ and $O_2$ in water than in air (Lange *et al.*, 2001).

**Evaporitic communities.** Cryptic communities are also found in salt deposits (Fig.1d) such as the cyanobacterial colonization of halite deposits in Baja, California (Rothschild *et al.*, 1994) and halite deposits in the Atacama Desert, Chile (Wierzchos *et al.*, 2006). Thus, even transient minerals produced by weathering and evaporative processes can host cryptic communities. The communities often form well-defined zones within the substrate, such as zones dominated by cyanobacteria and purple sulphur bacteria beneath them (Oren *et al.*, 1995). Evaporites containing halite require that their occupants are halophilic or at least halotolerant.

Communities in evaporitic deposits in extreme environments have been described. Similarly to endolithic organisms that inhabit the interior of rocks, the habitat can provide protection from environmental extremes. For example, cyanobacteria, principally *Chroococcidiopsis* morphotypes, inhabit evaporitic deposits in the Atacama desert (Wierzchos *et al.*, 2006). The organisms inhabit halite outcrops within the desert's hyperarid zone, achieving metabolic activity during periods of moisture availability, which is hypothesised to be assisted by halite deliquescence (Davilla *et al.*, 2008). Organisms within evaporites are not confined to extreme environments, however, and occur wherever evaporative conditions favour salt deposit formation. For example, the colonization of salt deposits in the Harz Mountains, Germany, by cyanobacteria has been investigated (Bioson *et al.*, 1994).

*Cryptic biota based on the various phases of water*

In addition to rock-forming minerals, the different phases of water, both on land and in water bodies, can host cryptic communities. Snow and ices can hide a photosynthetic biota beneath. The diversity of cryptic biota associated with these habitats is large. They include organisms covered by transient seasonal snow and ices and biota living permanently beneath snow and ice covers. The best-studied example of the latter biota is photosynthetic mats that inhabit the perennially ice-covered lakes of Antarctica (e.g., Hawes and Schwartz, 2001).

**Sea ice.** Sea ice typically has a brown layer at the bottom when viewed from beneath, or in section; this brown layer consists of (photosynthetic) diatoms and other biota in brine channels (Arrigo *et al.*, 1993, 1995, 1997; Lizotte, 2001; Thomas and Dieckmann, 2002). Sea ice occurs mainly in the Antarctic and shows great seasonal variability; at its maximum extent it covers $35 \times 10^6$ km$^2$, or 13% of the Earth's surface (Parkinson and Gloerson, 1993). Although ice can be transparent, in most natural environments heterogeneities in the ice structure and/or snow coverings make the biota spectrally hidden to observers above the material.

**Deep Chlorophyll Maximum.** Cryptic photosynthesis can occur in water bodies or oceans and could be relevant for planets with oceans or even covered entirely by oceans (Léger *et al.*, 2004). $O_2$-producing organisms at more than few metres depth of an ocean (Beckmann and Hense, 2007) or a lake are non-detectable in reflectance measurements due to the overlying water column that attenuates the reflected light from the organisms beneath (see Weston *et al.*, 2005; Xiu *et al.*, 2007). Photosynthetic pigments can only be remotely detected if they occur in the top few metres of the water body. Shorter wavelengths of PAR are absorbed by water to a smaller extent than is red and infra-red radiation so that "blue to green ratios" are used to detect chlorophyll (Platt and Sathyendrenath 1988; Falkowski and Raven, 2007; Xiu *et al.*, 2007).

The DCM phenomenon involves the occurrence of phytoplankton, typically measured as their chlorophyll, at a substantial depth in a large water body. This depth is usually more than 10 m and as great as 100 m or more in ocean waters with a low attenuation coefficient for PAR. DCMs can persist for periods of weeks to (at least) years, although with variations in their depth (Huisman *et al.* 2006).

The occurrence of DCMs involves phytoplankton organisms of neutral or negative buoyancy (i.e. sinking relative to the surrounding water) within a definite range of values of vertical mixing of the water column (Klausmeier and Litchman, 2001, Fennel and Boss, 2003, Hodges and Rudnik, 2004; Ruiz *et al.*, 2004; Raven and Waite, 2004; Huisman *et al.*, 2006; Beckman and Hense, 2007).

The formation of DCMs can be considered in the context of the inverse gradients of photosynthetically active radiation (PAR) and nutrients in the surface ocean, combined with sinking of phytoplankton that are denser than the surrounding water (Huisman *et al.*, 2006, Beckman and Hense, 2007). Attenuation of PAR by water, dissolved material and particles including organisms gives an exponential decay in PAR with depth (Falkowski and Raven, 2007). Phytoplankton-based productivity converts dissolved nutrients into particles, some of which sink and are biologically mineralised at depth, regenerating the nutrients. Return of the nutrients to the photic zone can be by permanent or seasonal upwelling (advection), or, over a larger area, by vertical turbulent diffusion (Huisman *et al.*, 2006).

A DCM forms when the supply of nutrients by turbulent diffusion is largely consumed in the DCM, with loss of phytoplankton and nutrients by sinking. The rate at which cells sink out of a DCM is an important factor in its dynamics and, in the limiting case, its steady state (Huisman *et al.*, 2006). Further to complicate the issue, Pilati and Wurtsbaugh (2003) have experimental evidence from a lake that persistence of a DCM requires the activities of zooplankton grazers.

The DCM represents an example of cryptic photosynthesis not obviously caused by extreme conditions, but rather by the interaction of the spatial constraints on the simultaneous availability of two resources, i.e. PAR and nutrients. To the extent that the phytoplankton under these conditions escape damage by high PAR (Raven, 1989) and/or UV at the surface, then their occurrence in a DCM could be regarded as means of avoiding extreme conditions. This situation may have been the case earlier in Earth's history between the time at which oxygenic photosynthesis originated and sufficient oxygen built up in the atmosphere to produced an effective atmospheric UV screen.

These examples show that cryptic photosynthesis can occur in many terrestrial and aquatic planetary environments; all lithologies can harbour cryptic photosynthesis (igneous, sedimentary or metamorphic), as can salts and minerals generated by evaporative processes, and environments containing liquid or frozen forms of water.

**THE PRODUCTION OF THE OXYGEN BIOSIGNATURE**
Can a cryptic biota produce an oxygen atmospheric signature? In this section we explore whether a cryptic world could produce a level of oxygen in the atmosphere that is detectable with future space based missions. The extent to which a cryptic photosynthetic biosphere could produce an oxygen biosignature on a given planet depends on a diversity of factors that are poorly quantified on Earth and unknown for extrasolar planets (Catling *et al.*, 2005). They include, *inter alia*: 1) the size of the planet, which will influence the extent of plate tectonics and hydrogen escape, 2) the area covered by the communities and the proportion of different cryptic biota, which will influence total oxygen production, 3) the quantity of abiotic reductants produced which must be oxidised before a net oxygen accumulation can occur, 4) the area of continents, which will influence carbon turnover and extent of long-term carbon sinks, 5) the flux and spectral quality of light from the host star, which will influence photosynthetic productivity (Wolstencroft and Raven, 2006), and 6) the sinks for reduced carbon, which will influence net oxygen accumulation. Oxygen accumulation in the atmosphere relies on there being a sink for the carbon reduced from $CO_2$, such as sequestration in the deep ocean or lithosphere.

Nevertheless, one can make a crude estimate of oxygen production using published data on cryptic photosynthesis. Cryptic photosynthetic communities show huge variability in their productivity. In Table 1 we have compiled a range of data from different communities and estimated the $O_2$ production of the communities based on published carbon uptake data. On land cryptic biota generally have lower productivity than non-cryptic biota and this probably reflects their frequent association with extreme environments. For example, Cockell and Stokes (2004, 2006) estimate high arctic hypolithic carbon uptake of 0.8 g $C/m^2$/year which can be compared to the estimates published by Raven (1995) of 0.9 μmol $CO_2/m^2$/s – equivalent to 380 g $C/m^2$/year – for a mat of filamentous cyanobacteria. This two order of magnitude difference would be consistent with the low light levels and cold temperatures that the hypoliths must endure on the underside of polar rocks and the transient summer activity that was assumed in the calculations (Cockell and Stokes, 2006). Johnston and Vestal (1991) concluded that Antarctic cryptoendolithic communities are about two orders of magnitude less productive than the open oceans.

The potential $O_2$ production by planetary-scale cryptic biota can be crudely calculated. Some workers have attempted to estimate the annual productivity of communities taking into account nanoclimate data acquired in the field. The mean net photosynthetic uptake of Antarctic cryptoendoliths is estimated to be 606 mg $C/m^2$/y (Friedmann *et al.*, 1993). Assuming a classic stoichiometric production of $O_2$ for $CO_2$ taken up in photosynthesis, this is equivalent to the production of 2.2 g $O_2/m^2$/y. If these communities covered the entire planetary surface then their annual $O_2$ output could be ~1.1 x $10^{15}$ g/y. The quantity of oxygen in the atmosphere today is ~1.5 x $10^{18}$ kg. Therefore, in theory these communities could produce the total quantity of $O_2$ found in the present atmosphere in about 1.3 Myr. Not all organic carbon is buried to cause net $O_2$ atmospheric accumulation. If 0.1% of the net carbon taken up by the endoliths is eventually buried as is assumed for deep sea sediments (Hedges, 2002), then it would take approximately 1.3 Gyr to accumulate today's oxygen concentration. However, generating this concentration is not required to produce a detectable oxygen biosignature. Kaltenegger *et al.* calculate that the lowest concentration of oxygen that could be detected in a planetary atmosphere is $10^{-3}$ PAL (Present Atmospheric Level) for a resolution of 70 in the visible and 25 in the mid IR (Kaltenegger *et al.* 2007). Thus, oxygen accumulation to detectable levels could occur much more rapidly.

There are obvious simplifications in this calculation. As Table 1 shows, the polar cryptoendolithic case is conservative and many cryptic habitats have much greater productivity. There is a trend towards increasing productivity in temperate cryptoendoliths and cryptic biota generally. At the same time, this calculation is obviously grossly simplistic as reductants are constantly being produced, such as in volcanism (and they would be at high concentrations in young atmospheres). This factor will reduce $O_2$ accumulation as it probably did on the early Earth. Finally, as listed at the beginning of this section, there are other factors that influence net oxygen accumulation in the atmosphere. However, the calculation shows that cryptic terrestrial photosynthesis could, under the right conditions, produce sufficiently large quantities of oxygen over geological time periods to produce an oxygen biosignature.

Similarly, in the oceans, cryptic photosynthesis exhibits smaller productivity than non-cryptic biota, but values may still be high enough to produce oxygen accumulation over geologic time periods. For example, sea ice communities are estimated to assimilate 0.063-0.070 Pg carbon in photosynthesis per year (Lizotte, 2001) in contrast to phytoplankton in this zone, which contribute about 1.3 Pg carbon per year. For comparison, the total marine primary production today (almost all from phytoplankton) has been estimated at about 50 Pg carbon per year (Field *et al.*, 1998), though higher values for the marine component have been suggested by del Giorgio and Williams (2005; see also Whittacker, 1975). This suggests that the cryptic photosynthesis by sea ice diatoms accounts for about 0.1 % of global marine productivity whereas potentially detectable phytoplankton in the sea ice zone account for more than 2% of global marine productivity.

For the DCM, Weston *et al.* (2005) found that it accounted for 37% of total annual new (i.e. potentially exportable) production in the summer-stratified North Sea. While the spatial and temporal extent of DCMs in the ocean is variable, these data suggest that DCMs contribute a significant amount to global primary productivity in the oceans, as in lakes (Moll and Stoermer, 1982). Table 1 shows that, where DCMs occur, they account for 5-90% of the primary productivity in the water column (total productivity, not just the 'new' exportable production).

The requirement for net oxygen accumulation is a carbon sink. On land, estimates of carbon sequestration by cryptic photosynthesis are unreliable since the long-term fate of fixed carbon from these communities has generally not been studied. The most important long-term sink for land-produced carbon is ocean sediment burial after riverine and Aeolian input (Hedges, 1992). Another plausible long-term sink for carbon is in freshwater sedimentary basins and subsequent burial, in a similar manner to coal formation. With respect to polar environments, where many of the cryptic biota discussed in this paper are located, permafrost is a potential long-term carbon sink (Zimov *et al.*, 2006). The permafrost of the Arctic is estimated to hold ~400 Gt of carbon and can act as a sink over mega-year time scales. On a cold planet with predominantly frozen conditions carbon could be leached from cryptic biota into the active zone to subsequently freeze and allow for net $O_2$ accumulation. Friedmann *et al.* (1993) investigated the fate of fixed carbon from Antarctic cryptoendoliths and calculated that about 50% of the carbon is lost into the environment without being respired. This carbon could become available for storage.

In the case of marine environment it seems likely that carbon from sea-ice communities or a DCM could be buried in deep-ocean sedimentary deposits in analogy to the Earth, allowing for net $O_2$ accumulation. We have been unable to find data on the life-time and ultimate sedimentation of the organic carbon produced in sea ice, but the DCM is better studied. Primary productivity in DCMs depends on input of nutrients by eddy diffusion (see Huisman *et al.* 2006). This productivity involving nutrient input from outside the zone where photosynthesis occurs is termed 'new production' by oceanographers and limnologists, in contrast to 'regenerated (or recycled) production' where remineralisation occurs in close proximity to the primary producers. 'New production' also includes surface productivity based on upwelled water, atmospheric inputs of combined (non-$N_2$) nitrogen, iron and phosphorus and biological nitrogen fixation. All forms of new production can be exported to deeper waters (e.g. Weston *et al.* 2008). Weston *et al.* (2008) indicated a very important role for the DCM in export production in the shallow southern North Sea. However, long-term net organic carbon burial depends on net input of combined nitrogen, phosphorus and iron and from biological nitrogen fixation in the water body. DCMs are typically remote from nutrient inputs from rivers, are 10 - ≥100 m vertically from the site of atmospheric nutrient inputs, and are often too deep for the most of the photosynthetic organisms that can carry out the energy-intensive process of biological nitrogen fixation (Sañudo-Wilhelmy *et al.* 2001; Letelier *et al.* 2004). This means that they seem to ultimately depend on the export component of surface primary productivity elsewhere in the ocean. However, on a planet whose oceanic primary productivity is all at 10 - ≥ 100 m and where there is no competition for nutrient from surface-dwelling primary producers there could be export production and burial of carbon from the DCM.

**RADIATIVE TRANSFER MODELS OF CRYPTIC PHOTOSYNTHESIS**

In this section we address two questions: 1) What would the spectral signature of cryptic photosynthesis worlds look like? and, 2) What is the range of a variety of cryptic world spectra, shown by eight different substrates for cryptic biota found on Earth ?

We calculate the model spectra for a variety of such cryptic world habitats assuming in our model that one biota found on Earth (see Table 1) dominates and compare the results to current Earth (for ease of comparison we keep the oxygen concentration at the same level for a cryptic world).

**Model description - Radiative transfer model to generate spectra of Earth and planets where the predominant photosynthetic output is cryptic.** The radiative transfer model we use to generate the spectra of cryptic planets and Earth is based on a code that was originally developed to calculate the Earth's infrared transmission spectrum (Stier and Traub, 1978), later extended to include infrared emission as well as visible transmission, and has since been used extensively for analyzing high resolution Fourier Transform spectra from ongoing stratospheric balloon-based observations to study the photochemistry and transport of the Earth's stratosphere (e.g., Jucks *et al.*, 1998). Our line-by-line radiative transfer code has also been used for numerous planetary disk modelling studies, both for theoretical work (e.g. Des Marais *et al.*, 2000; Traub and Jucks 2002) and for fitting observed Earthshine spectra (Woolf *et al.*, 2002; Turnbull *et al.*, 2006) and infrared data (Kaltenegger *et al.*, 2007), and finally for calculating the visible and infrared spectrum of the Earth over geological time (Kaltenegger *et al.*, 2007). The model provides an excellent fit to observed reflection and emission data sets (Woolf *et al.*, 2002; Turnbull *et al.*, 2006; Christensen and Pearl, 1997).

We use a simple geometrical model in which the spherical planet is modelled with a plane parallel atmosphere and a single angle of incidence and reflection. This angle is selected to give the best analytical approximation to the integrated-Earth air mass factor of two for a nominal illumination (quadrature); the zenith angle of this ray is 60 deg.

We model Earth's and cryptic world's reflection spectra using Earth's spectroscopically most significant molecules at Earth's present atmospheric pressure ($H_2O$, $O_3$, $O_2$, $CH_4$, $CO_2$, $CFC_{11}$, $CFC_{12}$, $NO_2$, $HNO_3$, $N_2$ and $N_2O$). We divide the atmosphere into 30 thin layers from 0 to 100 km altitude. The spectrum is calculated at very high spectral resolution, with several points per line width, where the line shapes and widths are computed using Doppler and pressure broadening on a line-by-line basis, for each layer in the model atmosphere. We include opacities due to Rayleigh scattering for each layer in the model atmosphere. We assume that the different reflecting layers (one surface and three cloud layers) can be approximated by 4 parallel streams that are co-added to produce the overall spectrum. All streams traverse the same molecular atmosphere, but each stream reflects from a different surface. For example, the first stream reflects from the planet's surface at 0 km altitude, the second and third stream reflect from a cloud layer with a top at an adjustable height (1 km and 6 km), and the fourth stream reflects from a cloud at a high altitude (12 km).

We adopt an overall 60% cloud cover factor for our Earth and planet models, with the relative proportions of cloud at each altitude being set to be consistent with the Earthshine data (Woolf *et al*., 2002; Turnbull *et al*., 2006). The proportional factors are 40% of the total at 1 km, 40% at 6 km, and 20% high cloud at 12 km for today's Earth. Most surface features like ice or sand show very small or very smooth continuous reflectivity changes with wavelength (albedo data from ASTER 1999, USGS 2003) (see Kaltenegger, 2007 for details). In our models we assign 70% of the planetary surface as ocean, 2% as coast, and 28% as land. The land surface for present-day Earth consists of 30% grass, 30% trees, 9% granite, 9% basalt, 15% snow and 7% sand. For the modelled cryptic worlds, the fraction of Earth's surface that is covered by vegetation (60%) is assumed to be made out of materials associated with cryptic photosynthesis habitats. These individual surfaces for the present-day Earth as well as cryptic photosynthesis habitats (no vegetation) are co-added percentage-wise to model a realistic Earth surface (this part of the model is shown in case a). For our final models the spectra from the surfaces and clouds are co-added with the indicated weights (case b) (see Fig.2).

The overall high-resolution spectrum is calculated, and smeared to lower resolution, like that proposed for missions such as Darwin/TPF. For reference and further explanation concerning the code, the reader is referred to our calculation of a complete set of molecular constituent spectra, for a wide range of mixing ratios, for the present Earth pressure-temperature profile, for the visible and thermal infrared, in Des Marais *et al*. (2002), and for the Earth over geological time, in Kaltenegger *et al*. (2007).

**Results.** Fig. 2 summarises our results and show the expected spectral signatures of worlds harbouring cryptic photosynthetic biota. Fig. 2 shows reflection spectra from 0.5 to 2 μm (the visible to near-IR region) for eight different cryptic photosynthesis worlds compared to present-day Earth with Earth cloud coverage for a disk averaged view. To highlight the difference in the overall detectable spectra, we model two scenarios for every cryptic world: a) we replace the fraction of Earth's surface that exhibits vegetation with the habitat of the chosen habitat and b) we assume that there were no clouds on that planet or their contribution could be averaged out of the spectrum. Case b) is chosen to explore the difference of the surface features of cryptic worlds in detail, but is currently considered unrealistic for a remote observation of a planet. Note that there is a factor of 5 in scale between the plots a) including and b) excluding clouds, demonstrating the high albedo of clouds in comparison to surface albedos. Clouds reduce any surface features (e.g. the red edge) considerably.

To remotely detect potential life on other planets with space mission in preparation, we explore the absorption features in its spectrum and whether any of these features cannot be explained without a biological source to produce them. On Earth the high oxygen concentration in our atmosphere is a sign of biological activity, especially in combination with methane.

As signs of life in themselves $H_2O$ and $CO_2$ are secondary in importance because although they are raw materials for life, they are not unambiguous indicators of its presence. Both existed a long time in Earth's atmosphere before life appeared. The oxygen and ozone absorption features in the visible and thermal infrared respectively could have been used to indicate the presence of photosynthetic biological activity on Earth anytime during the past 50% of the age of the solar system, while the red edge reflection feature evolved only during the most recent 10% of the age of the solar system (Kaltenegger et al., 2007). Oxygen is likely to be biogenic when found in combination with water if the planet's temperature is moderate and a runway greenhouse stage– like proposed for early Venus – can be excluded (see e.g. Selsis et al., 2002; Segura et al., 2007).

In the visible to near-infrared one can see increasingly strong $H_2O$ bands at 0.73 μm, 0.82 μm, 0.95 μm, and 1.14 μm. The strongest $O_2$ feature is the saturated Fraunhofer A-band at 0.76 μm. A weaker feature at 0.69 μm can not be seen with low resolution. $O_3$ has a broad feature, the Chappius band, which appears as a broad triangular dip in the middle of the visible spectrum from about 0.45 μm to 0.74 μm that can also be seen. $CO_2$ has negligible visible features at present abundance and accounts for the weak 1.06 μm feature.

Figure 2 illustrates that cryptic worlds would show surface reflectance spectral signatures in the 0.4 to 2 μm range entirely explained by abiotic materials for a variety of biota found on Earth (see Table 1), but are habitable worlds..

**CONCLUSIONS**
On a hypothetical planet where surface vegetation never developed, or surface conditions are more extreme than on present-day Earth, or a high impact rate favours cryptic photosynthesis, or where the complete surface is ocean-covered, a substantial amount of $O_2$ production could occur within cryptic photosynthetic communities. On such a planet a surface biosignature such as the Vegetation Red Edge would never develop. The reflection surface spectrum of such extrasolar worlds would be dominated by standard mineral and water reflections. Prior to the rise of oxygen in the Earth's atmosphere and the formation of an ozone shield, the initial production of oxygen would have been accomplished by a photosynthetic biota protected from UV radiation that was essentially cryptic. Marine cryptic photosynthesis would have been most important as net carbon burial would likely have been more effective in the oceans than on land, and therefore would have made a greater contribution to atmospheric oxygenation. From this standpoint, the Earth during the onset of the Great Oxygenation Event ~2.4 Ga ago until the appearance of a red edge ~0.45 Ga ago may have exhibited a cryptic photosynthesis signature.

**ACKNOWLEDGEMENTS**
Charles Cockell gratefully acknowledges support from the Science and Technology Facilities Council (STFC). John Raven gratefully acknowledges discussion with Professor Lyn Jones on the evolution of infrared reflectivity in photosynthetic organism on land. Lisa Kaltenegger gratefully acknowledges support from the Origin of Life Initiative and the NASA Astrobiology Institute. The University of Dundee is a registered Scottish Charity, No. SC015096.

Table 1. The forms of cryptic photosynthetic biota on the Earth. Examples of estimated oxygen production from different environments and in different cryptic biota are provided and the references from which the estimates were obtained[1,2]

| Cryptic photosynthetic biota | Example | Comments on potential planetary scale ubiquity | References to other examples | Examples of potential oxygen production |
| --- | --- | --- | --- | --- |
| Chasmoendoliths (organisms within cracks) | Antarctic granites (De Los Rios et al., 2005) | Any substrate with cracks induced by freeze-thaw, aqueous or aeolian weathering, impact cratering, thermal stress, tectonism, etc. can provide a habitat (all substrates in Fig. 2) | Broady, 1981; Cockell et al., 2002 | Not measured, but presumed to be similar to cryptoendoliths |
| Cryptoendoliths (organisms within rock interstices) | Dry Valley Antarctic quartzitic sandstones (Friedmann, 1982) | Growth within interstices implies high porosity and permeability, usually associated with sedimentary lithologies (e.g., quartz sandstones, limestones). However, processes such as impact cratering may render substrates more suitable for colonisation | Friedmann, 1980; Bell, 1993; Bell and Sommerfield, 1987; Wessels and Büdel, 1995; Kurtz and Netoff, 2001; Saiz-Jimenez et al., 2001; Cockell et al., 2002 | 0.0004-0.016 g $O_2/m^2/y$ (Vestal, 1988 - Antarctic polar desert)[a]; 8.8-44 g $O_2/m^2/y$ (Nienow et al. 1988 - Antarctic polar desert)[a,3], 0.0005-0.002 g $O_2/m^2/y$ (Johnston and Vestal, 1991 - Antarctic polar desert)[a,4]; 4 g $O_2/m^2/y$ (Friedmann et al., 1993 - Antarctic polar desert); 0.078-0.58 g $O_2/m^2/y$ (Tretiach and Pecchiari, 1995 – North-East Italy); 0.02 g $O_2/m^2/y$ (Matthes-Sears et al., 1997 – Ontario, Canada)[b,5] |
| Hypoliths (organisms on the underside of rocks) | Polar desert rocky environments (this paper; Cockell and Stokes, 2004, 2006) | Hypoliths generally require translucent rocks (e.g., quartz) for transmission of PAR. However, freeze-thaw in polar deserts can allow cryptic photosynthesis under any substrate e.g., igneous, basalt; metamorphic, granite; and sedimentary rocks (all substrates in Fig. 2) | Berner and Evenari, 1978; Büdel and Wessels, 1991; Büdel et al., 2004; Smith et al., 2000; Schlesinger et al., 2003; Warren-Rhodes et al., 2006 | 2.9 g $O_2/m^2/y$ (Cockell and Stokes, 2004, 2006 – Arctic polar desert)[a]; 378 g $O_2/m^2/y$ (Schlesinger et al., 2003 – Mojave Desert, USA) |
| Euendoliths (organisms that actively bore into rocks) | A diversity of coastal carbonates (Schneider and Le Campion-Alsumard, 1999) | Mainly carbonates, implying sedimentary environments, but borings into volcanic glass and feldspars are reported | Golubic et al., 1981; Fisk et al., 1998; Hoppert et al., 2004; Smits, 2006 | Not measured |
| Soil and desert crusts (organisms living within sands and soils) | Desert crusts ubiquitous in Chinese deserts (Chen et al., 2007) | Requires grains of sand within which organisms can form a crust, implying aqueous or aeolian weathering of rock | Campbell, 1979; Eldridge and Greene, 1994; Danin et al., 1998; Dickson 2000; Garcia-Pichel et al., 2001; Hu and Liu, 2003; Johansen, 2004 | 1526 g $O_2/m^2$/year (Lange et al., 1992 – Negev Desert, Israel)[b]; 6900-10,753 g $O_2/m^2$/year (Brostoff et al., 2002 – Mojave Desert, USA)[b]; 950-2640 g $O_2/m^2$/year (Garcia-Pichel and Belnap, 2008 – Utah, USA)[c] |
| Sand/soil cryptic communities (organisms living beneath surface soil layer, not necessarily forming a crust) | Mats living beneath soil surfaces in Yellowstone National Park, USA (Rothschild, 1994) | Requires grains of soil/sand within which organisms grow, implying aqueous or aeolian weathering of rock | Potentially any phototrophs living in soils (e.g., Thomas and Dougill, 2006; Rahmonov and Paitek, 2007) | 55 g $O_2/m^2$/year (Rothschild, 1994 – Wyoming, USA)[a] |
| Evaporitic communities (organisms within salt crusts) | Organisms inhabiting halite mounds. Atacama Desert, Chile (Wierzchos et al., 2006) | Requires regions with high evaporation rates and/or low precipitation rates to allow for formation of salt crusts | Rothschild et al., 1994; Douglas, 2004; Douglas and Yang, 2002; Spear et al., 2003 | 0.00025 g $O_2/m^2$/year (Rothschild et al., 1994 – California, USA)[a] |
| Snow and ice habitats (where substrate is sufficiently heterogeneous to block the reflected light signature from the biota beneath), including biota in sea ice. | Organisms under perennially ice-covered lakes in Antarctica (Tanabe et al., 2008) Organisms in brine channels within sea ice (Arrigo et al., 1997; Lizotte, 2001; Thomas and Dieckmann, 2002) | Provided the snow and ice is not so thick as to reduce light to below the minimum required for photosynthesis and temperatures remain above the minimum for photosynthesis (Kappen et al., 1996), most snow and ice covers can harbour cryptic biota | Hawes and Schwartz, 2001; Priscu et al., 1998; Squyres et al., 1991 | >365 g $O_2/m^2$/year (Arrigo et al., 1993, 1995, 1997 - Antarctic sea ice, assuming summer productivity continues throughout the year, i.e. at least twice the real value)[a] |
| Aquatic communities | Maximum zones (deep chlorophyll maxima) of phytoplankton (Cox et al., 1982; Strass | Potentially any water body more than ~ 10 m deep, granted appropriate mixing | Benthic photosynthetic organisms below a few metres depth (Littler et | 28-57 g $O_2/m^2$/year (MacKey et al., 1997 – western Equatorial Pacific, DCM 46-75% of total water column |

| | and Woods, 1991; Pollehne et al., 1993; MacKey et al., 1997; Klausmeier and Litchman, 2001; Fennel and Boss, 2003; Hodges and Rudnik, 2004; Lee and Whitledge, 2005; Weston et al., 2005; Huisman et al.,2006; Beckmann and Hense, 2007; Hanson et al., 2007; Xiu et al., 2007) | characteristics. Usually co-exists with phytoplankton nearer the surface which might be susceptible to remote sensing | al., 1986; Raven et al., 2000), or in stromatolites (Raven et al., 2008) | (TWC) production)[a], 10.7 g $O_2/m^2$/year (Lee and Whitledge 2005 – Canada Basin, Arctic Ocean open water, 80% of TWC)[d], 239.3 g $O_2/m^2$/year (Weston et al., 2005 – central North Sea, 58% of TWC during summer stratification)[a], 10-195 g $O_2/m^2$/year (Hanson et al.2007 – Leeuwin Current, subtropopical south-east Indian Ocean, 5-90% of TWC)[a] |

[1] These numbers are estimates. Many of the measurements are based on $^{14}$C-bicarbonate uptake for which there is an ambiguous separation between gross and net C uptake. Those estimates using gas exchange measurements could be an underestimate if rates of respiration were high during measurements. Some of the estimates are based on productivity measurements during productive periods of growth in the field or in laboratory settings (e.g., Rothschild and Giver, 2003; Schlesinger et al., 2003). We have extrapolated these to annual numbers but of course in the field there are seasonal variations with periods of lower productivity in winter months which are generally not reported and laboratory numbers may not reflect field productivity.

[2] For comparison Raven (1995) provides estimates for many microbial non-cryptic biota, e.g. 1248 g $O_2/m^2$/y for a mat of filamentous cyanobacteria, 4577 $O_2/m^2$/y for epilithic green algae with a $CO_2$ pump, 4993 $O_2/m^2$/y for liverworts with intercellular gas spaces. The open ocean is estimated to produce about 475 g $O_2/m^2$/y (Whittacker, 1975, Field et al., 1998, del Giorgio and Williams 2005))

[3] This is an overestimate. Nienow et al. express their figures as carbon uptake per algal cell. In this calculation we have assumed that an algal cell is 10 μm in diameter and they are packed edge to edge over a square meter. The biomass is obviously not this great in nature.

[4] This is based on carbon uptake into lipids which will be a fraction of total carbon uptake and this is therefore an underestimate.

[5] This may also contain an epilithic component

[a] = studies using $^{14}$C-bicarbonate uptake

[b] = gas exchange measurements

[c] = oxygen microelectrode measurements

[d] = studies using $^{13}$C-bicarbonate uptake

Fig. 1. Four examples of land-based cryptic photosynthetic communities. a) Hypoliths (inset, arrow) inhabit the underside of a wide diversity of substrates in the Canadian High Arctic in regions where rocks are sorted into 'patterned' ground, b) a cryptoendolithic lichen (arrow) inhabiting the interstices of sandstone in the Dry Valleys of the Antarctic, c) chasmoendoliths (arrow) inhabit an impact-fractured rock in the Haughton impact crater, Canadian High Arctic, d) endoevaporites inhabit a salt crust visible as pink pigmentation (arrow) (photo: Marli Bryant Miller).

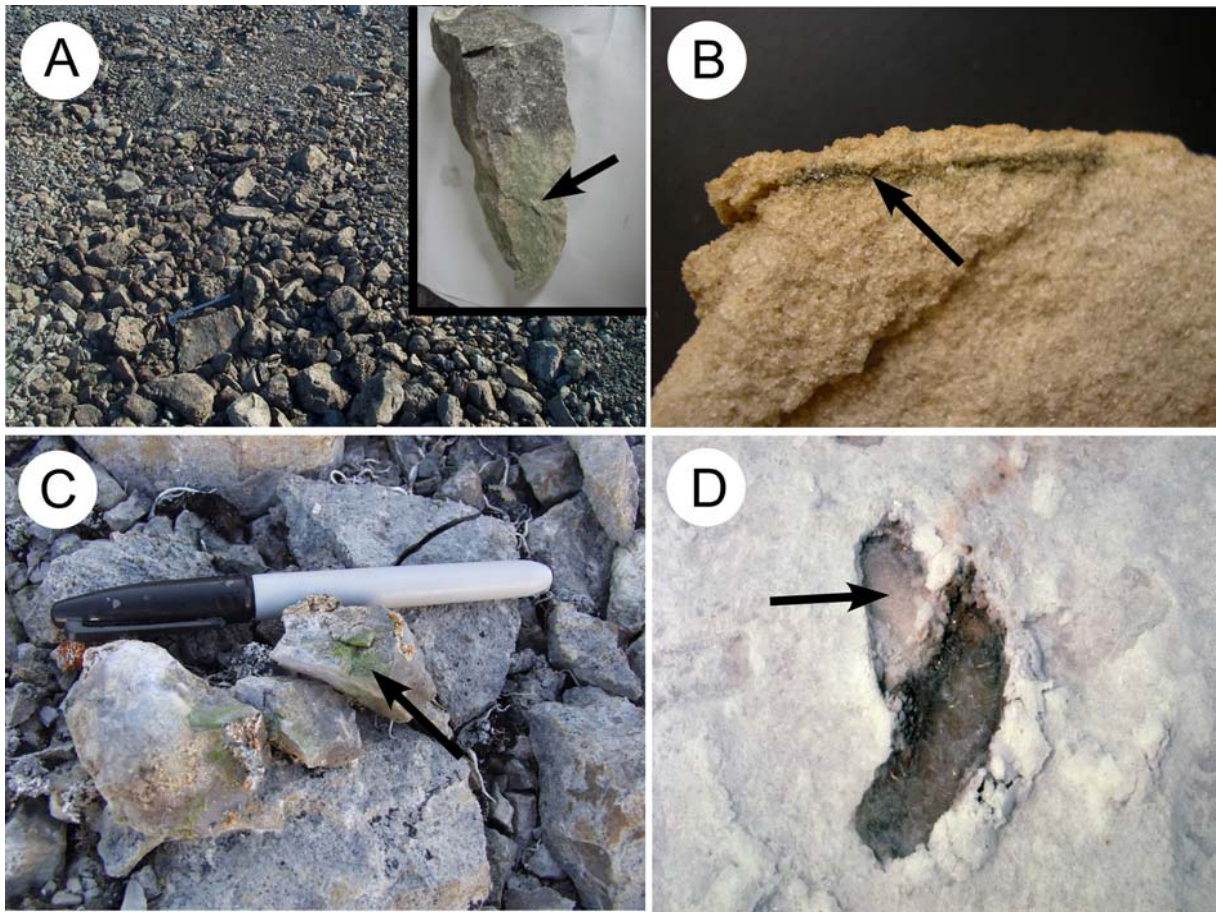

Fig. 2. Calculated reflection spectra from 0.4 to 2 μm (the visible to near-IR region) for different cryptic photosynthesis habitats (left: snow, salt, sea) (right: sand, basalt, granite) compared to present-day Earth with clouds (case b, upper panel) and with no clouds (case a, lower panel) for a disk averaged view. We assume here that the surface area covered by vegetation on current Earth is replaced by the habitat of the chosen cryptic biota. Substrates represent typical habitats for different cryptic biota (see Table 1).

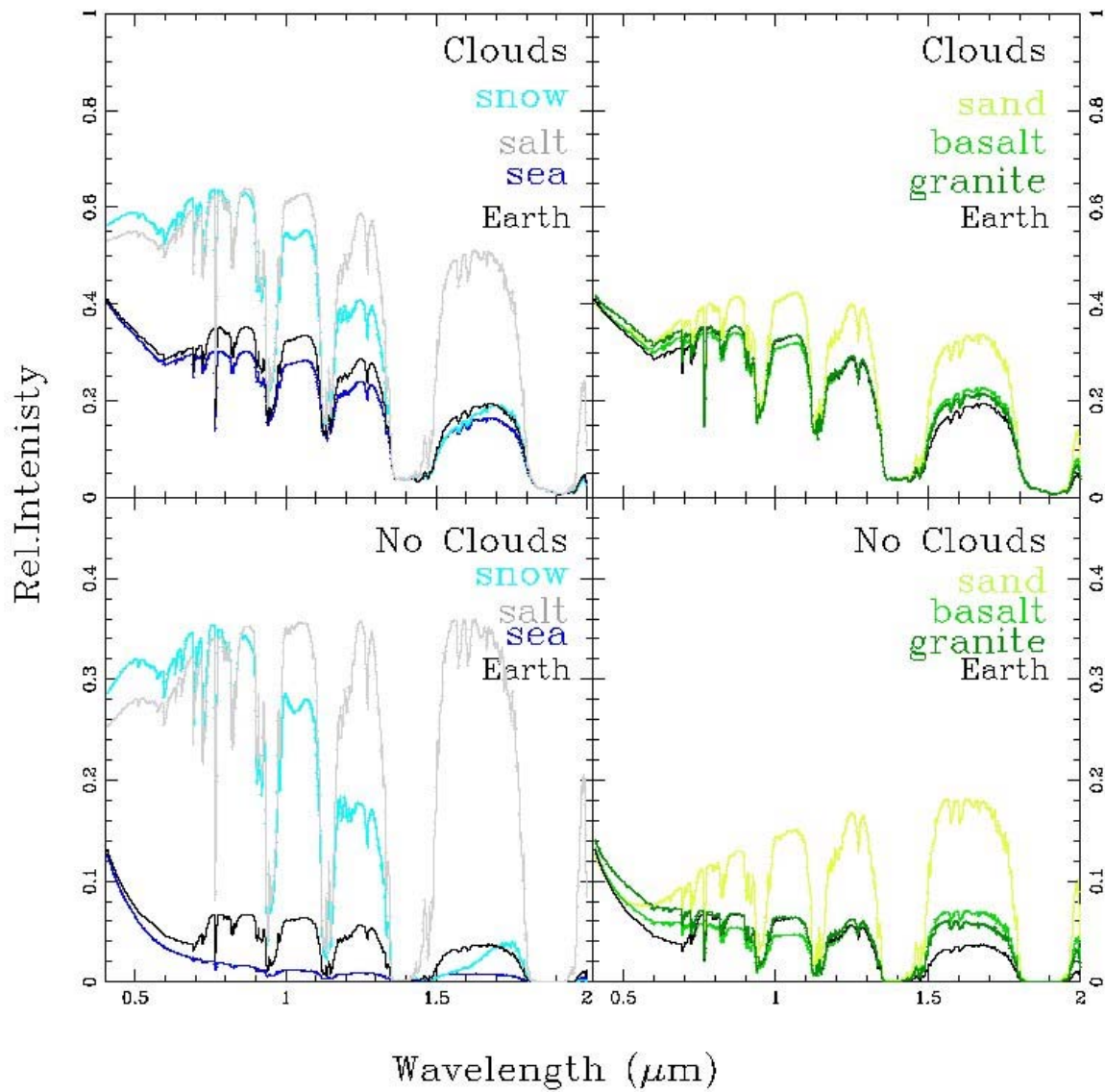